# Model-based Approach for Analyzing Prevalence of Nuclear Cataracts in Elderly Residents

**Sachiko Kodera[1], Akimasa Hirata[1,2], Fumiaki Miura[1], Essam A. Rashed[1,3], Natsuko Hatsusaka[4], Naoki Yamamoto[4], Eri Kubo[4], Hiroshi Sasaki[4]**

1. Department of Electrical and Mechanical Engineering, Nagoya Institute of Technology, Nagoya 466-8555, Japan

2. Center of Biomedical Physics and Information Technology, Nagoya Institute of Technology, Nagoya 466-8555, Japan

3. Department of Mathematics, Faculty of Science, Suez Canal University, Ismailia 41522, Egypt

4. Department of Ophthalmology, Kanazawa Medical University, Ishikawa 920-0293, Japan

Corresponding Author

Akimasa Hirata

Department of Electrical and Mechanical Engineering, Nagoya Institute of Technology

Gokiso-cho, Showa-ku, Nagoya, Aichi, 466-8555, Japan.

E-mail: ahirata@nitech.ac.jp

**Abstract**

Recent epidemiological studies have hypothesized that the prevalence of cortical cataracts is closely related to ultraviolet radiation. However, the prevalence of nuclear cataracts is higher in elderly people in tropical areas than in temperate areas. The dominant factors inducing nuclear cataracts have been widely debated. In this study, the temperature increase in the lens due to exposure to ambient conditions was computationally quantified in subjects of 50–60 years of age in tropical and temperate areas, accounting for differences in thermoregulation. A thermoregulatory response model was extended to consider elderly people in tropical areas. The time course of lens temperature for different weather conditions in five cities in Asia was computed. The temperature was higher around the mid and posterior part of the lens, which coincides with the position of the nuclear cataract. The duration of higher temperatures in the lens varied, although the daily maximum temperatures were comparable. A strong correlation (adjusted $R^2 > 0.85$) was observed between the prevalence of nuclear cataract and the computed cumulative thermal dose in the lens. We propose the use of a cumulative thermal dose to assess the prevalence of nuclear cataracts. Cumulative wet-bulb globe temperature, a new metric computed from weather data, would be useful for practical assessment in different cities.

# 1. Introduction

The eye is highly sensitive to various environmental agents [1-3]. International standardization bodies set the limits for non-ionizing radiation, especially for wavelengths less than 1 mm, based on eye safety [4]. However, the mechanisms are variable, including corneal damage, photochemical reactions, and cataract formation [5, 6].

The role of environmental agents in cataractogenesis, which has been observed predominantly in elderly populations, has been a controversial and widely debated topic. Although extensive studies have been conducted, the exact mechanism of cataractogenesis is not yet fully understood. This could be attributed to a number of cofactors influencing cataractogenesis [7] and should be investigated separately and systematically. This study focused specifically on physical factors, among which ultraviolet (UV) radiation and ambient temperature exposure [2, 6, 8, 9] have received much attention.

The three most common types of cataracts are nuclear cataracts (NUCs), which lead to a gradual opacification of the nucleus of the lens; cortical cataracts, which involve the cortex from the periphery toward more central opacification; and posterior subcapsular cataracts, which give rise to distinct opacity on the posterior capsule [10]. Several studies have reported that UV radiation is a risk factor for cortical cataracts. Within the solar radiation spectrum [11], UV is an electromagnetic wave with a wavelength between 10 and 400 nm, corresponding to a penetration depth between 20 and 150 μm in the human skin [12]. A recent epidemiological study suggested that UV radiation is correlated with the prevalence of cortical cataracts [13].

Miranda argued that cataract incidence in tropical and subtropical zones showed a stronger correlation with ambient temperature than with solar radiation [14]. This study also suggested that NUCs are more prevalent in tropical and subtropical zones. Another aspect to be noted is that the prevalence of NUCs increased significantly in the elderly, particularly those living in tropical zones [15, 16]. According to a recent epidemiological study, self-reported cataracts were observed 4 times more frequently among people aged ≥ 70 years than among those aged 50–59 years (95% CI: 2.28－7.50) [17]. If cataractogenesis in the tropical zone is closely related to ambient temperature and/or solar radiation, the number of patients may increase in subtropical zones as climate change continues and aging societies persist [18].

Weather data from a measurement-based database can provide useful insights on ambient temperatures and support future research. Ambient temperature and relative humidity (RH) may influence body temperature. Solar radiation, specifically infrared radiation, should also be considered, and it is considered to vary over time. Energy deposition is concentrated around the cortical region of the skin and cornea because of the small penetration depths, and it then diffuses into deeper regions due to heat conduction [19, 20].

To the best of the authors' knowledge, cataractogenic exposure has not been quantified before. Thus, the internal physical quantity needs to be evaluated. It is difficult to evaluate lens temperature because, in addition to a complex heat transfer system around the eye, the thermophysiological response may change with age as well as location (i.e., tropical and temperate areas). In our previous studies, we modeled the thermoregulatory response in the elderly based on the measurement of human subjects [21, 22]. In addition, we clarified that the tropical test subject was mainly characterized by parameters corresponding to the countercurrent blood in different body parts and an approximately 20% increase in sweating rate [23].

If we can characterize NUC prevalence in different countries, the results may facilitate the development of future intervention strategies. In this study, we investigated the effect of lens temperature as a potential physical agent for NUC formation. First, we extend a conventional thermoregulatory response model of the test subject to elderly subjects in a tropical region. The temperature in the eye lens, particularly around the center, was computed based on exposure to ambient temperature and infrared radiation in adult males and elderly individuals who grew up in tropical zones. We introduced the cumulative thermal dose in the mid-part of the lens as a potential physical agent of NUC in the elderly population of different locations based on epidemiological studies conducted by our group [13, 16]. The corresponding ambient parameter was then correlated with the lens temperature for practical assessment.

## 2. Materials

*2.1. Data sources for epidemiological evaluation*

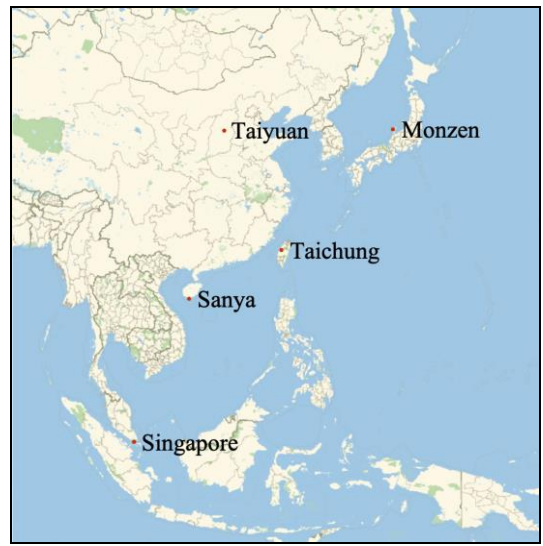

Figure 1. Map of east Asia showing the five target cities.

The prevalence of NUC has been measured and reported in previous studies from Kanazawa Medical University [13, 16]. The subjects comprised 2,637 residents aged 50 or over (average 62.4 ± 9.2 years) who participated in epidemiological surveys conducted in five cities (shown in Fig. 1; generated with Wolfram Mathematica 12.1) with varying UV irradiance and ambient temperatures (Table 1). The study was approved by the Institutional Review Board of Kanazawa Medical University and was conducted according to the principles of the Declaration of Helsinki; written informed consent was obtained from all participants.

The lens opacities of all participants were assessed by the same ophthalmologist using a slit-lamp microscope under maximum mydriasis. NUC was classified according to the WHO Cataract Grading System [10], and participants with grade NUC-1 and above in either the left or right eye were eligible for further analysis. Those with intraocular lenses in both eyes or unspecified cases were excluded from this study.

Table 1. Latitude and ambient conditions in the hottest month in five different cities (Sanya, Singapore, Taichung, Taiyuan, and Monzen).

| Cities | Sanya, China | Singapore | Taichung, Taiwan | Taiyuan, China | Monzen, Japan |
| --- | --- | --- | --- | --- | --- |
| Latitude [°] | 18.25 N | 1.29 N | 24.14 N | 37.87 N | 37.29 N |
| Hottest month | May | August | August | July | July |
| Highest temperature [°C] [24] | 33.9 | 33.7 | 32.6 | 34.7 | 32.3 |
| Humidity [%] | 64 | 49 | 64 | 30 | 61 |
| UV irradiance [mW/m$^2$] [25] | 211.8 | 255.9 | 189 | 113 | 110.8 |

Table 2. Number of subjects, demographic characteristics, and prevalence (%) of nuclear cataracts in the five evaluated cities (Sanya, Singapore, Taichung, Taiyuan, and Monzen).

| Cities | Sanya, China | Singapore | Taichung, Taiwan | Taiyuan, China | Monzen, Japan |
|---|---|---|---|---|---|
| Total Number | 513 | 473 | 555 | 671 | 425 |
| Gender (M / F) | (190 / 323) | (216 / 257) | (227 / 328) | (250 / 421) | (165 / 260) |
| Mean age ± SD | 61.2 ± 10.8 | 61.6 ± 7.4 | 60.9 ± 8.1 | 60.6 ± 9.0 | 69.8 ± 7.2 |
| Aged 50–59 (50s) | | | | | |
| Prevalence [%] | 21.6 | 6.4 | 2.3 | 1.4 | 0 |
| (95% CI) | (16.6–26.6) | (3.1–9.8) | (0.6–4.1) | (0.2–2.7) | |
| Num. of subjects | 264 | 202 | 293 | 350 | 35 |
| Mean age ± SD | 52.5 ± 2.7 | 54.9 ± 2.8 | 54.7 ± 2.9 | 53.3 ± 2.5 | 57.6 ± 1.1 |
| Aged 60–69 (60s) | | | | | |
| Prevalence [%] | 74.8 | 38.9 | 13.4 | 7.3 | 0.6 |
| (95% CI) | (66.8–82.7) | (32.2–45.6) | (8.1–18.7) | (3.5–11.2) | (0.0–3.2) |
| Num. of subjects | 115 | 203 | 157 | 177 | 177 |
| Mean age ± SD | 63.2 ± 2.8 | 64.1 ± 2.8 | 64.1 ± 2.9 | 64.1 ± 2.8 | 65.1 ± 2.9 |
| Physiological responses model | Tropical | Tropical | Temperate | Temperate | Temperate |

*2.2. Weather Data*

The ambient temperature and RH for the five cities in 2019 were obtained from the National Oceanic and Atmospheric Administration's (NOAA's) National Data Center Climate Data Online (NNDC CDO) [24]. The global solar radiation and altitude values were obtained from Weather Spark [26].

## 3. Model and Methods

*3.1. Anatomical human model*

A Japanese male model based on magnetic resonance images was used as a reference in this study [27]. The height and weight of this model were 1.73 m and 65 kg, respectively, which are close to the mean measurements of Japanese males. This model was segmented into 51 anatomical regions, such as skin, bone, muscle, and fat, including six tissues around the eye (e.g., lens, vitreous, iris, cornea, and vitreous). The resolution of the original model was 2 mm. However, this resolution is not sufficient to represent the eye anatomy (e.g., ciliary body, choroid, etc.). Thus, a digital super-resolution method was used to improve model resolution to 1 mm using a smoothing algorithm [28] (241,302,240 voxels). The eye and its surrounding tissues were then manually corrected to avoid modeling artifacts in the original full-body model.

Figure 2 illustrates the anatomical human model, illustrates body parts used to determine blood temperature, and presents the enlarged anatomical figures around the eye, which were derived from the whole-body model. The body-part model is conceptually identical to a compartment model (for example, a cylinder in [24]).

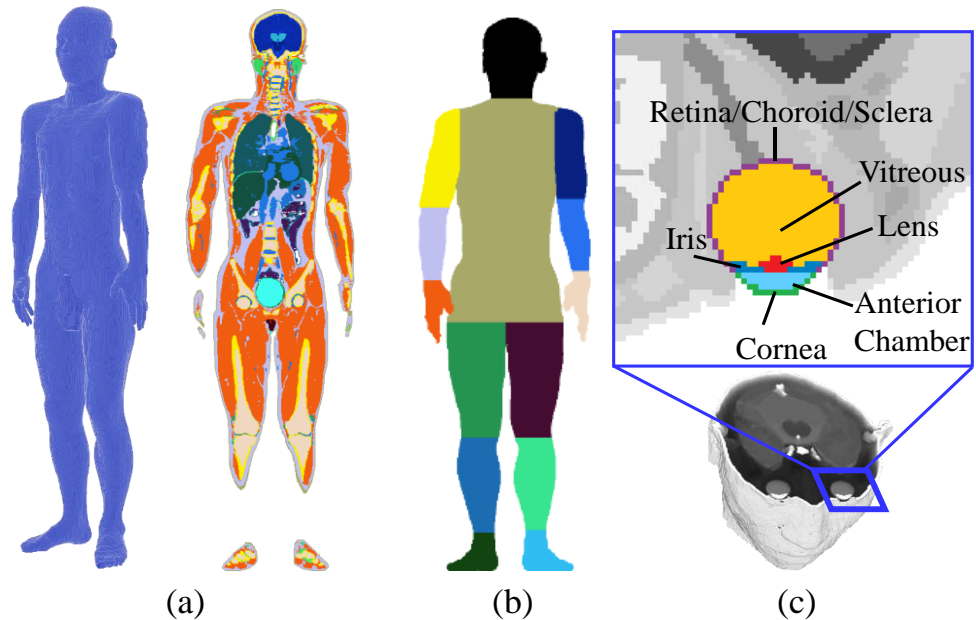

Figure 2 (a) Anatomical model of a male adult, (b) corresponding body part model used to compute blood temperature and evaporation loss, and (c) eye anatomy (derived from the whole-body model). The whole-body model resolution is 1 mm.

*3.2 Thermal analysis*

A detailed explanation and validation of our in-house computational code is provided in our previous study [29]. The thermodynamics in biological tissues (Sec. 3.2.1) and the thermoregulatory response (Sec. 3.2.3) were jointly computed. After computing temperature in the tissue of the anatomical model, the blood temperature was computed (Sec. 3.2.2) in each part of the model. To compute the tissue temperature with solar radiation, the power deposition distribution computed by the electromagnetic computation (Sec. 3.3) was used to analyze the tissue temperature computation (Sec. 3.2.1). The computational code was implemented on an SX-ACE supercomputer [30] with vectorization and parallelization, similar to our previous study [29, 31]. The computational time to determine the 24-h exposure to thermal load was 36 min on a supercomputer with 64 nodes (estimated as 550 h on a workstation with 24 nodes).

*3.2.1. Modeling based on bioheat equation*

Pennes' bioheat transfer equation [32] is often applied to follow the human body temperature [33, 34] in the time domain. A generalized equation considering the time variation of thermoregulation and core temperature [22] can be expressed as:

$$C(\mathbf{r})\rho(\mathbf{r})\frac{\partial T(\mathbf{r},t)}{\partial t} = \nabla \cdot \left(K(\mathbf{r})\nabla T(\mathbf{r},t)\right) + M(\mathbf{r},t) + \sigma(\mathbf{r})E^2(\mathbf{r},t) - B(\mathbf{r},t)\left(T(\mathbf{r},t) - T_B(m,t)\right), \quad (1)$$

where $\mathbf{r}$ and $t$ denote the three-dimensional position vector and time, $T$ and $T_B$ are the tissue temperature and the blood temperature in each body part ($m$ = 1, …, 14), respectively, and $m$ represents the body parts shown in Fig. 2 (b); $C$, $K$, $B$, and $\rho$ denote the specific heat, thermal conductivity, blood perfusion rate, and mass density of the tissue, respectively; $M$ is the metabolic heat generation that depends on human activity; and $\sigma$ and $E$ denote the conductivity of the tissue and the internal electric field caused by solar radiation, respectively.

The boundary conditions between the air and tissue were determined using Eq. (1):

$$-K(\mathbf{r})\frac{\partial T(\mathbf{r},t)}{\partial n} = H(\mathbf{r},t) \cdot (T(\mathbf{r},t) - T_a(t)) + EV(\mathbf{r},t), \tag{2}$$

where $T_a$ is the ambient temperature, $EV$ is the evaporative heat loss, and $H$ is the heat transfer coefficient between air and tissue, which includes the convective and radiative heat losses [35]. The heat transfer coefficient is given as follows:

$$\begin{aligned}
H(\mathbf{r},t) &= (H_c(\mathbf{r},t) + H_r(\mathbf{r},t))/R_{sf}, \\
H_c(\mathbf{r},t) &= \left\{ a_{nat} (T_{sf}(t) - T_a(t))^{1/2} + a_{nat} \cdot v + a_{mix} \right\}^{1/2}, \\
H_r(\mathbf{r},t) &= \sigma_s \psi \varepsilon_{sf} \varepsilon_{sr} \cdot \left\{ (T_{sf}(t) + 273)^2 + (T_a(t) + 273)^2 \right\} \cdot \left\{ (T_{sf}(t) + 273) + (T_a(t) + 273) \right\},
\end{aligned} \tag{3}$$

where $H_c$ and $H_r$ denote the convective and radiative heat transfer coefficients, respectively, as also reported in [36]. The variable $v$ denotes the wind velocity, and $T_{sf}$ denotes the average body surface temperature. $a_{nat}$ (=4.63), $a_{frc}$ (=199.74), and $a_{mix}$ (=−9.8) are the corresponding regression coefficients. $\sigma_s$ (=5.67×10$^{-8}$ W/m$^2$/K$^4$) is the Stefan-Boltzmann constant, $\psi$ (= 0.794) is the corresponding view factor, and $\varepsilon_{sf}$ (=0.99) and $\varepsilon_{sr}$ (=0.93) are the emissivity of the body surface and surrounding indoor space, respectively [37]. The surface area of the voxel-based model with 1 mm resolution is approximately 1.4 times [38] as large as that of an actual human estimated from [39]. The heat transfer coefficient is adjusted by the ratio between the actual and voxelized body surface area ($R_{sf}$=1.38). The temperature of the internal air contained in the lung/esophagus was set to the average surrounding tissue and ambient temperature [40], and the heat transfer coefficient was treated as in Eq. (3).

Equation (1), subjected to boundary condition (2), was solved for the three-dimensional whole-body model (Fig. 2 (a)) with the finite-difference time-domain (FDTD) method. The discretization is a uniform rectangular grid with a resolution of 1 mm, which coincides with the model resolution. Note that an isolated eye model is not used here for the thermal conduction between the eye and surrounding tissues, considering the power absorption of solar radiation in the region surrounding the eye (especially in the skin) [41] (see Secs. 3.3 and 3.4).

In addition, the coefficient between the cornea and air is presented in Appendix B. The specific heat, blood perfusion rate, heat conductivity, and basal metabolism were identical to those presented in our previous study [42] as listed in Appendix A. The blood perfusion rate varied due to an increase in the tissue temperature, as reported in [43, 44].

*3.2.2 Modeling of Vasodilation Heat Loss*

Vasodilation modeling according to temperature elevation above a certain level [44] was coupled with bioheat modeling (Eq. (1)). Here, the modeling in [40] was applied. The blood perfusion rate in the skin through vasodilation is expressed as follows:

$$B_{skin}(\mathbf{r},t) = \left(B_{0,skin}(\mathbf{r}) + F_{HB}\Delta T_H(t) + F_{SB}\Delta T_S(t)\right) \cdot 2^{\Delta T(\mathbf{r},t)/6}, \qquad (4)$$

where $B_{0, skin}$ (=3,680 W/m³/°C) is the skin basal blood perfusion rate, $F_{HB}$ (=17,500 W/m³/°C²) and $F_{SB}$ (=1,100 W/m³/°C²) denote the coefficients of the signals from the hypothalamus and skin, respectively [40], and $\Delta T_H$ and $\Delta T_S$ denote the average temperature rise in the hypothalamus and average value of the skin, respectively.

The blood perfusion rate in all tissues except for the skin is expressed as follows:

$$\begin{aligned} B(\mathbf{r},t) &= B_0(\mathbf{r}) & T(\mathbf{r},t) < 39\,°C, \\ B(\mathbf{r},t) &= B_0(\mathbf{r})\{1 + S_B(T(\mathbf{r},t) - 39)\} & 39\,°C \leq T(\mathbf{r},t) < 44\,°C, \\ B(\mathbf{r},t) &= B_0(\mathbf{r})(1 + 5S_B) & 44\,°C \leq T(\mathbf{r},t), \end{aligned} \quad (5)$$

where $B_0$ denotes the basal blood perfusion rate in each tissue and $S_B$ (=0.8 °C⁻¹) denotes a coefficient defining changes in the blood perfusion characteristics.

*3.2.3. Computation of blood temperature*

The variation in blood temperature was computed considering arterial and venous temperatures. The countercurrent heat exchange was considered to obtain a more realistic distribution of the arterial blood temperature, rather than assuming a constant arterial blood temperature across all body parts equivalent to the temperature of the central blood pool. For the body parts defined in Fig. 2 (b), the blood temperature was computed. The arterial blood temperature variation was based on [45] and is given by the following equation:

$$T_{bla,m}(t) = \frac{T_{blp}(t) \sum_{\mathbf{r}}^{voxels} B(\mathbf{r},t)V(\mathbf{r})}{h_{x,m} + \sum_{\mathbf{r}}^{voxels} B(\mathbf{r},t)V(\mathbf{r})} + \frac{h_{x,m} \sum_{\mathbf{r}}^{voxels} T(\mathbf{r},t)B(\mathbf{r},t)V(\mathbf{r})}{\sum_{\mathbf{r}}^{voxels} B(\mathbf{r},t)V(\mathbf{r}) \left( h_{x,m} + \sum_{\mathbf{r}}^{voxels} B(\mathbf{r},t)V(\mathbf{r}) \right)}, \quad (6)$$

The venous temperature can be expressed by the following equation, based on the assumption that capillary blood reaches an equilibrium with the surrounding tissue:

$$T_{blv,m} = \frac{\sum_{\mathbf{r}}^{voxels} T(\mathbf{r},t)B(\mathbf{r},t)V(\mathbf{r})}{\sum_{\mathbf{r}}^{voxels} B(\mathbf{r},t)V(\mathbf{r})}, \quad (7)$$

where $T_{bla,m}$ and $T_{blv,m}$ denote the arterial and venous temperatures in each body part $m$, respectively. The blood perfusion rate, $\mathbf{B}$, was computed from Eqs. (4) and (5); the parameters $V$ and $h_{x,m}$ are the tissue volume and coefficients corresponding to the counter-current heat exchange, respectively. The value of $T_{blp}$ is the central blood pool temperature and is defined as the heat exchange by integrated blood perfusion and temperature throughout the body as follows:

$$T_{blp} = \frac{\sum_{m} \left( \dfrac{\sum_{\mathbf{r}}^{voxels} B(\mathbf{r},t) \cdot V(\mathbf{r}) \cdot \sum_{\mathbf{r}}^{voxels} T(\mathbf{r},t) \cdot B(\mathbf{r},t) \cdot V(\mathbf{r})}{h_{x,m} + \sum_{\mathbf{r}}^{voxels} B(\mathbf{r},t) \cdot V(\mathbf{r})} \right)}{\sum_{m} \left( \dfrac{\left( \sum_{\mathbf{r}}^{voxels} B(\mathbf{r},t) \cdot V(\mathbf{r}) \right)^2}{h_{x,m} + \sum_{\mathbf{r}}^{voxels} B(\mathbf{r},t) \cdot V(\mathbf{r})} \right)}, \quad (8)$$

The blood temperatures in the head and torso were assumed to be the same due to their high blood perfusion rate [29]. The values of V, $h_x$, and $m$ are given in Appendix A.

*3.2.4 Modeling of evaporative heat loss in adult and elderly subjects*

Several thermoregulatory response models have been proposed [13, 25]. Our algorithm for sweating is based on the one presented in [43] and is extended to elderly subjects in the temperate zone [46] and younger adults in the tropical zone [23]. The thermoregulatory response modeling for the elderly (> 65 years) is mainly characterized by a decline in the sweating rate and the retarded sweating response, which was expressed by considering the heat sensitivity in the skin [21, 22]. The thermoregulatory response modeling for younger adults in tropical zones is characterized by heat adaptability according to different birthplace climates [23]. Both thermoregulation response models have been validated in these studies. No study has attempted to express the thermoregulation of the elderly living in a tropical area based on two modeling studies. We propose a combined thermoregulatory response model for the elderly living in tropical zones.

The evaporative heat loss from the skin is given as follows:

$$\begin{aligned} EV(\mathbf{r},t) &= \min\{SW(\mathbf{r},t)\cdot 40.6/S,\ EV_{max}(t)\}, \\ EV_{max}(t) &= 2.2\cdot h_c f_{pcl}\left(P_S(t) - \varphi_\varepsilon P_A(t)\right), \end{aligned} \qquad (9)$$

where $SW$ is the sweating rate [g/min] defined in Eqs. (10) and (11) (see below), $S$ [m$^2$] is the total surface area of the human body, and 40.6 is a conversion coefficient [J·min/g/s]. The maximum evaporative heat loss, $EV_{max}$, on the skin depends on ambient conditions. The convective heat transfer coefficient is represented by $h_c$; $P_S$ and $P_A$ are the saturated water vapor pressures at the temperature of the skin and at the ambient air temperature, respectively; $\varphi_e$ is the RH of the ambient air; and $f_{pcl}$ is the permeation efficiency factor of clothing, which is affected by the wind speed. For simplicity, $f_{pcl}$ is assumed to be 0.8, corresponding to a lightly clothed body [47, 48].

The sweating rate *SW* was assumed to depend on the temperature of the skin and the body core (approximated as hypothalamus), according to the following equation [43]:

$$SW(\mathbf{r},t) = \gamma(\mathbf{r}) \cdot \begin{bmatrix} \{\alpha_{11} \tanh(\beta_{11} \Delta T_{S+}(t) - \beta_{10}) + \alpha_{10}\} \Delta T_{S+}(t) \\ + \{\alpha_{21} \tanh(\beta_{21} \Delta T_{H+}(t) - \beta_{20}) + \alpha_{20}\} \Delta T_{H+}(t) \end{bmatrix} + PI, \quad (10)$$

where $\Delta T_S$ and $\Delta T_H$ are the temperature increases of the skin averaged over the body and hypothalamus temperatures, respectively. The insensible water loss (*PI*) of younger adults and elderly was 0.71 g/min, based on the weight and height [49]. Note that the insensible water loss is minor factor in the condition of the scenario. The multiplier $\gamma(\mathbf{r})$ denotes the dependence of the sweating rate on body parts [43]. The coefficients $\alpha$ and $\beta$ were estimated for the average sweating rate based on the measurements in [44] and those obtained from our previous study [50].

Different modeling methods for thermoregulation for the elderly have been proposed in different studies [21, 22, 51, 52]. The modeling for the elderly is based on the methodology in our previous study [21]. The maximum sweating rates for all body parts except for the limbs were comparable between younger adults and the elderly [53-56]. Equation (10) was modified to model the elderly as follows [46]:

$$SW(\mathbf{r},t) = \chi(\mathbf{r})\gamma(\mathbf{r}) \cdot \left[ \left\{ \alpha_{11} \tanh\left(\beta_{11}(\Delta T_S(t) - T_{S,dec})_+ - \beta_{10}\right) + \alpha_{10} \right\} \cdot (\Delta T_S(t) - T_{S,dec})_+ \right.$$
$$\left. + \left\{ \alpha_{21} \tanh\left(\beta_{21}\Delta T_{H+}(t) - \beta_{20}\right) + \alpha_{20} \right\} \Delta T_{H+}(t) \right] + PI, \quad (11)$$

where $\chi(\mathbf{r})$ (=0.6 in the legs) denotes the coefficients related to the reduction of sweating in the limbs. The parameter $T_{S,dec}$ (= 1.5 °C) denotes the threshold for the sweating response in the elderly, which represents an aging-related deterioration in the thermal sensitivity of the skin [21]. A decline in sweating rate due to heat sensitivity in the hypothalamus was observed in subjects older than 74 years but was not significant for subjects around 65 years of age [22].

The total active sweat glands of the tropical subjects were approximately 1.2 (1.11–1.23) times higher than that of subjects from temperate climates due to long-term acclimation [57]. The measured sweating rate of tropical subjects was 1.2–1.5 times higher than that of temperate subjects when exposed to heat [58, 59]. In this study, the sweating rate of people from tropical regions was modeled by multiplying Eqs. (10) and (11) with 1.2, which was validated in our previous study [23]. The model of thermoregulation for the elderly living in a tropical zone was simply combined these characteristics as will be discussed in the Discussion section. An additional parametric study considering corneal temperature is provided in the Appendix B.

*3.3 Solar radiation*

The computation of power deposition in tissues due to solar absorption was the same as in our previous study [11,12]. The FDTD method for Maxwell's equation was used. An in-house FDTD code was used, which was validated via intercomparison [60]. The computational domain was truncated with a twelve-layered perfectly-matched layer as the absorbing boundary condition [35]. The dielectric properties of the biological tissues were obtained from the database in [36]. The power absorption was calculated from the Joule loss ($\sigma E^2$ in Eq. (1)); the electric field in the tissue is computed.

The penetration depth of the infrared radiation is approximately 4 mm. Thus, as an incident wave, the electromagnetic field at 6 GHz is considered for simplicity. Additional discussion regarding the validity of simulating solar absorption can be found in [11]. The transmittance of the clothes for infrared radiation was chosen as 0.32 [37]. The test subject is assumed to be half-sleeved shirt and long pants.

*3.4 Exposure scenarios*

The initial temperature distribution in the whole-body model was computed at the steady state without a heat load. The distribution was obtained from the bioheat equation (1) subject to the boundary condition (2) with the assumption that the body core temperature is 37.0 °C and

37.3 °C with a basal metabolism for temperate and tropical test subjects, respectively. Core temperature and thermoregulatory responses were considered unchanged without a heat load. Note that diurnal change of body core temperature normally fluctuates ± 0.5 °C, following circadian rhythms [61]. Also, this fluctuation depends on the season and age [62]. The core temperature of 37.0 °C approximately coincides with the measured results of 7:45 am in [62] for the temperate area.

Here, the effect of the thermoregulatory response, age (50s for adults and 60s for elderly), and solar radiation on the lens temperature was evaluated. First, as a fundamental study, an ambient temperature of 35 °C and RH of 50% were considered.

Subsequently, realistic cases were considered to simulate core temperature changes with the time course of ambient temperatures and RH. However, it is almost impossible to conduct a computational analysis for all months considering the computational costs. First, the hottest month, whose daily maximum temperature is greater than 31 °C, was determined from the weather data. The mean ambient conditions over five hottest days were derived using the arithmetic average, as shown in Fig. 3. Then, the temperature change over the whole-body were computed for 24 hours considering time series of ambient condition in Fig. 3. Even for heat-related illness, such canonical exposure scenarios worked well [63].

The other heat load is solar radiation, which generates heat due to electromagnetic power absorption. Considering variations in human behavior, the thermal doses were computed with and without solar radiation in each city. It is difficult to consider all aspects of human behavior and duration spent outside and inside the building. In addition, the effect of solar radiation on the human body depends on the relative direction of the subject to the sun. Further, the reflection from the ground depends on the material (e.g., soil, stone-paved, etc.). It is unrealistic to assume that the solar radiation will be incident on the eye directly (normally) for several hours. The mean value of solar radiation averaged over the azimuth direction was applied to the human body. The reflection coefficient was assumed to be 0.15 considering normal ground, for simplicity. Detailed modeling and validation for power absorption in the model due to solar radiation is reported in our previous studies [29, 64].

Based on the climate classification, the test subjects in Singapore and Sanya (China) were considered as the tropical population, whereas those in Taichung (Taiwan), Taiyuan (China), and Monzen (Japan) were considered to be in temperate areas. The population model in the tropical and temperate areas mentioned above was then applied to these cities based on climate.

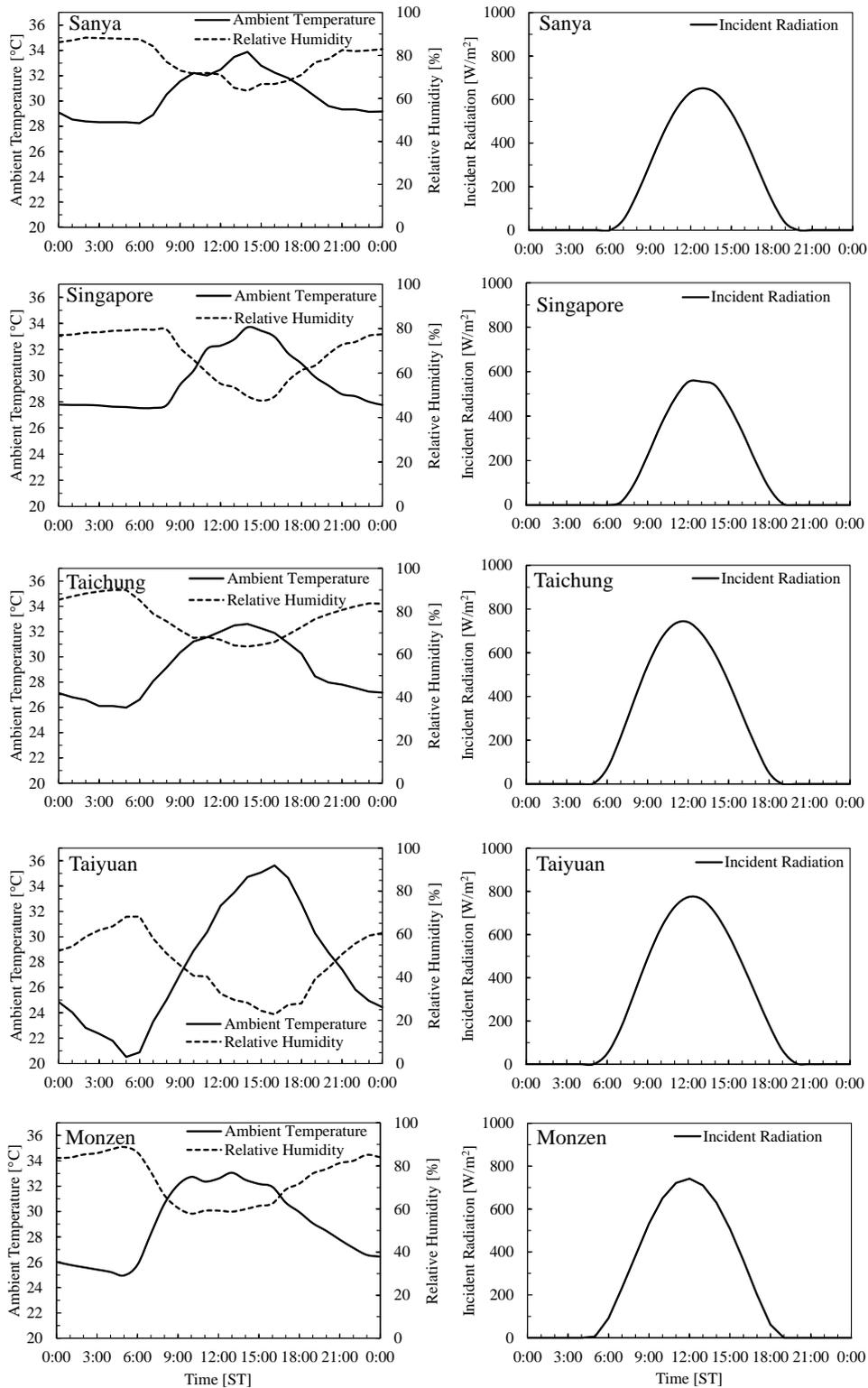

Figure 3. Ambient temperature, RH, and solar radiation averaged over the hottest month for different cities covered in this study.

*3.5. Evaluation metrics*

To evaluate the effect of different factors on the prevalence of NUC, the ambient factor was considered: i) monthly averaged ambient temperature over the hottest month, ii) yearly averaged ambient temperature, iii) monthly averaged wet bulb globe temperature (WBGT) averaged over the hottest month, and iv) UV radiation. In addition, to test the hypothesis that NUC is related to heat load, the daily peak lens temperature was considered v) with and vi) without solar radiation.

In addition to the ambient factor, the cumulative thermal dose in the lens was considered to affect the difference in prevalence according to age. Cumulative ambient factors and lens temperature were also considered, as shown below.

$$\int \max\left\{\left(T_a(t) - T_{a,th}\right), 0\right\} dt, \qquad (12)$$

where $T_{a,th}$ is the threshold value of the WBGT. A threshold ambient temperature of 28 °C was selected, which corresponds to the thermoneutral condition in a lightly clothed adult. The WBGT threshold of 24 °C was selected, which corresponds to an ambient temperature of 28 °C for an RH of 50%:

$$\int \max\left\{\left(WBGT(t) - WBGT_{th}\right), 0\right\} dt \qquad (13)$$

The cumulative thermal dose, which is defined as the period at which the lens temperature is higher than a certain temperature multiplied by the temperature increase (e.g., increases from 37 °C) is introduced as a metric to associate with the increase in prevalence of NUC.

There was an implicit assumption that the non-NUC population in their 50s is completely healthy, and a thermal dose over 10 years is essential to cause cataract formation.

$$\int \frac{T_{a,\max}(i)}{T_{a,\max,year}} \int \max\left\{(T_{lens}(t) - T_{lens,th}), 0\right\} dt\, di, \qquad (14)$$

where $T_{lens}(t)$ and $T_{lens,th}$ denote the time course of the computed lens temperature throughout the day (24 h) and the threshold of lens temperature, respectively. As the computed temperature was the mean value of the five hottest days, a scaling factor was applied to estimate the 1-year cumulative value. $T_{a,max}(i)$ and $T_{a,max,year}$ denote the daily ($i = 1,\ldots, 365$) and yearly maximum ambient temperature, respectively, which were estimated in the same way as the weighting of the WBGT. Note that the 10-year period considered here is the difference in the data of subjects aged 50–59 and 60–69, which is the cumulative thermal dose during this period.

The thermal damage in biological tissues is characterized by the cumulative equivalent minutes at 43 °C ($CEM_{43}$) [65]. However, this metric is generally applicable to a relatively short exposure to heat, and thus was not directly applied to this study. The threshold value of $CEM_{43}$ in the lens has been reported to be much lower than that in other tissues, a few times to one order magnitude smaller than the remaining eye tissue (for rabbits) [65]. For simplicity and to account for the high uncertainty of its behavior, we defined here the cumulative heat dose as the product of the increase in lens temperature and the number of days.

## 4. Computational Results

### 4.1. Fundamental characteristics of lens temperature for heat exposure

The lens temperatures in different test subjects (age and climate) were evaluated at an ambient temperature of 35 °C and RH of 50%. The time required to reach a steady state was 30 min (90% saturation). As shown in Fig. 4, the difference in the lens temperature between the temperate and tropical zones was 0.3 °C, which is mainly attributable to the difference in the core temperatures. In the elderly, the core temperature increased due to the decline in heat sensitivity of the skin and hypothalamus, resulting in a lowered sweating rate. The core temperature difference of elderly subjects aged around 65 years was 0.3 °C higher than that of subjects in their 50s. This suggests that the central part of the lens temperature is approximately 0.6 °C higher in the elderly living in tropical regions than in the adults living in the temperate regions. For typical exposure scenarios, the temperature in the lens may reach 37.0–37.5 °C or higher. A higher temperature rise appeared around the mid to posterior part of the lens, which coincides with the location of the NUC (see Fig. 4).

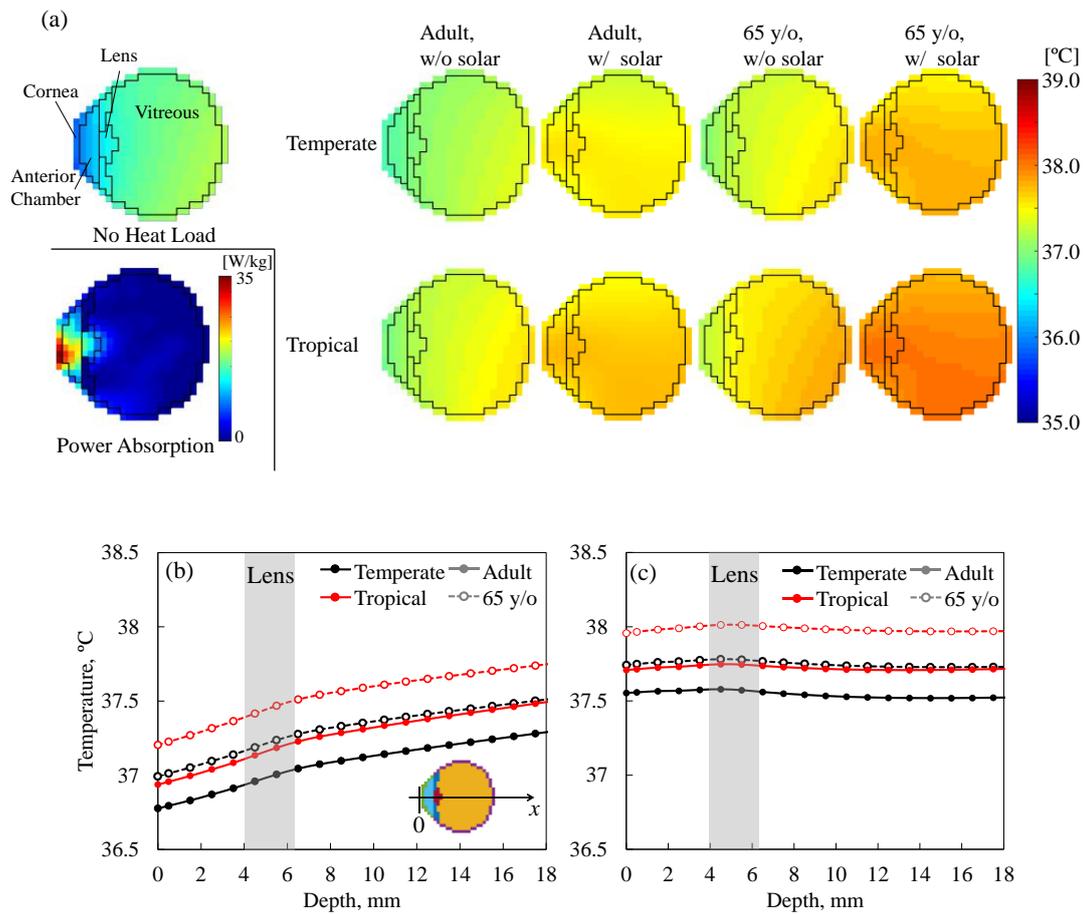

Figure 4. (a) Temperature distribution and power absorption in the horizontal cross section to stay in the ambient temperature of 35 °C and RH of 50% at steady state: (a) thermoneutral condition, (b) without and (c) with solar radiation (150 W/m$^2$), along the central axis.

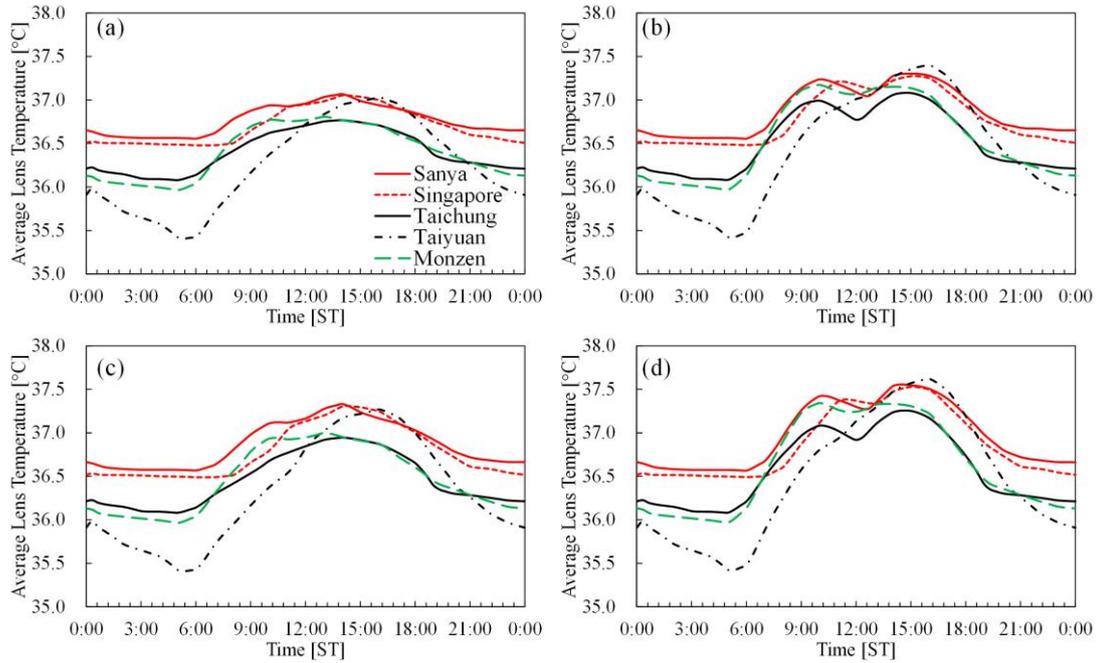

Figure 5. Time course of average lens temperature in test subjects of (a, b) 50s (adult model) and (c, d) 60s (65 y/o model) for ambient conditions of five different cities (in Fig. 2): (a, c) and (b, d) are the results for cases without and with solar radiation.

*4.2. Lens Temperature Rise in Populations of Different Cities*

Figure 5 shows the time course of mean lens temperature rise in test subjects in their 50s and 60s for ambient conditions of different cities. The simulation was conducted for a 24-h period. As shown in Fig. 5, the lens temperature was the highest in Sanya, which was comparable to Singapore, followed by Taiyuan (China), Taichung (Taiwan), and Monzen (Japan). It should be noted that the duration for which the lens temperature was higher than a certain level (e.g., 37.0 °C) differed by city.

*4.3. Correlation between NUC prevalence and heat load*

Figure 6 shows the relationship of ambient temperature (averaged (a) for a year and (b) the hottest month), (c) average WBGT for a year, (d) average UV radiation for a year, and peak lens temperature ((e) without and (f) with solar radiation) with the prevalence of NUC. As shown in Fig. 6 (a–d), a mild correlation ($0.20 <$ adjusted $R^2 < 0.53$) was observed between the prevalence of NUC and parameters related to the ambient conditions. As shown in Fig. 6 (e) and (f), the prevalence can be approximated with lens temperature as a linear function, where the slopes for subjects in their 50s and 60s were found to be different. For all parameters, the *p*-value was larger than 0.05. The difference in the prevalence between subjects aged 50–59 and 60–69 led to the hypothesis that the cumulative effect should be considered.

The prevalence difference (those of subjects aged 60–69 minus those aged 50–59) was estimated with the cumulative heat dose, as shown in Fig. 7. The threshold of the lens temperature was set at 37.0 °C. The temperature rise due to modest activity was also assumed (METS=2-3). The correlation is better for the cumulative thermal dose (adjusted $R^2$=0.920 and

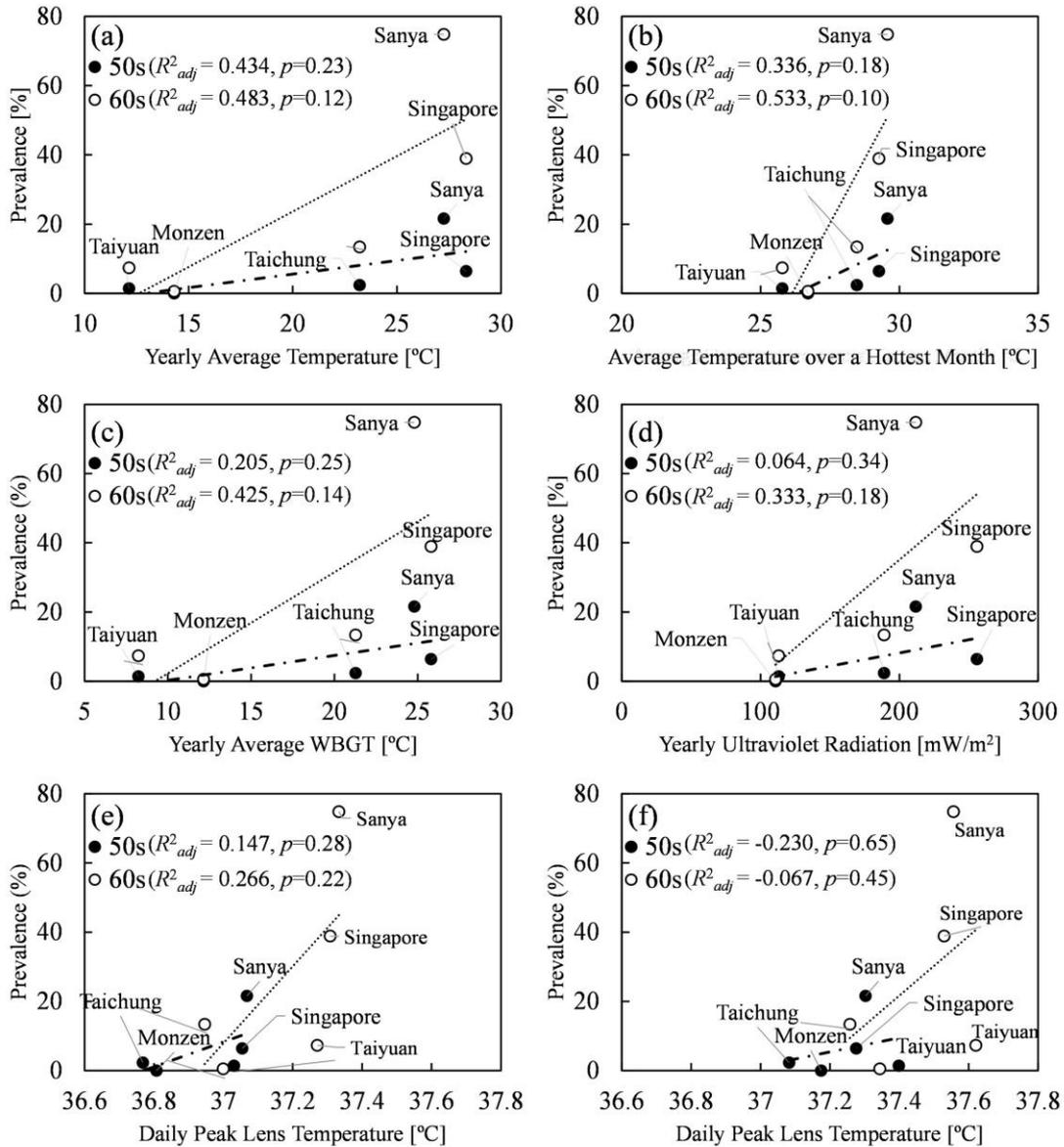

Figure 6. Correlation of NUC prevalence with ambient temperature averaged (a) over an year and (b) for the hottest month, (c) average WBGT for one year, (d) average UV radiation for an year, and daily peak lens temperature (e) without and (f) with solar radiation.

$R^2$=0.850 with and without solar radiation, respectively) rather than cumulative values of ambient conditions. Except for cumulative UV, the p-value was smaller than 0.05.

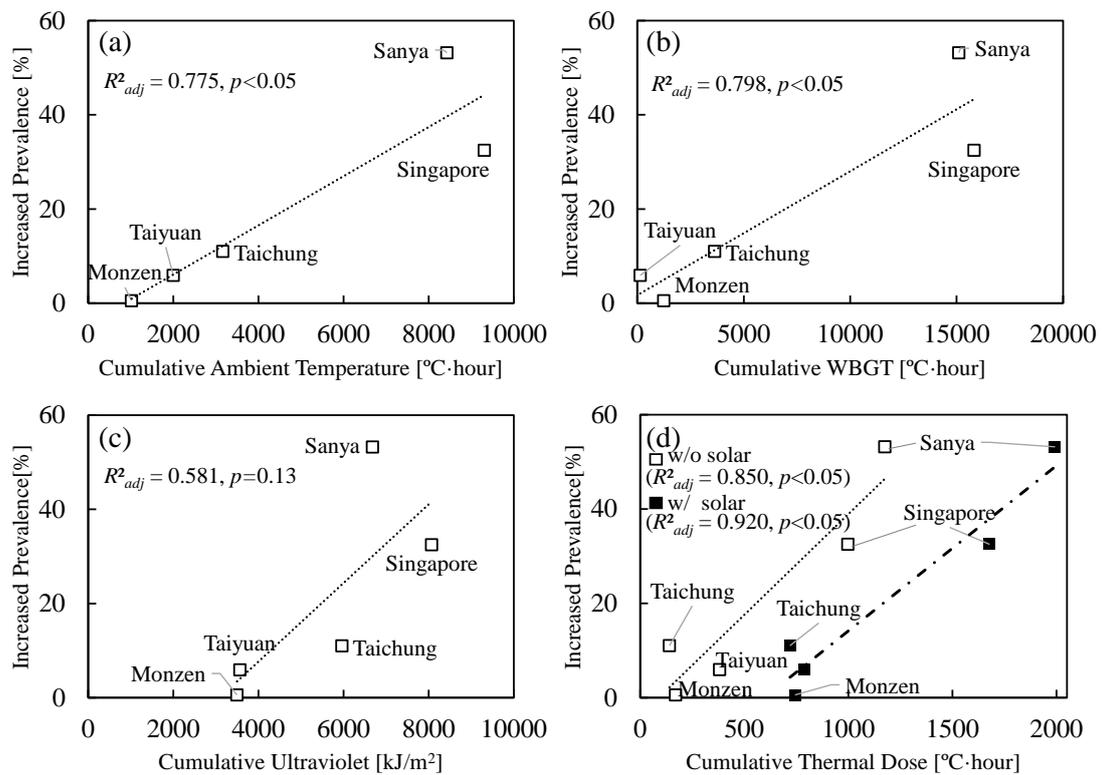

Figure 7. Relation of increased NUC prevalence with (a) cumulative ambient temperature, (b) WBGT, (c) ultraviolet, and (d) thermal dose in lens.

## 5. Discussion

For test subjects in conditions where the ambient temperature is 35 °C and RH is 50%, the core temperature difference of the elderly around 65 years (60s) was 0.3 °C higher than that of healthy adults (50s). As often reported in the field of heat-related illnesses, this can be attributed to a decline in the sweating capacity of the elderly or low evaporative heat loss [66]. This is also different for populations in temperate and tropical areas. Our computational results may explain the observation that core temperature elevates by 0.3 °C in the afternoon [62]. Once the core temperature elevates, the vasodilatation (eq. (4)) and sweating (eq. (2)) are activated, lowering core temperature rise or to maintain core temperature around 37.0 °C [67]. We computationally demonstrated that this increase in body core temperature resulted in an increase in the temperature of the lens nucleus via blood circulation. Owing to the finite blood perfusion rate, the eye temperature increase is not identical to that of the blood. It is crucial to note that the position of high temperature in the lens appears around the mid to posterior region of the nucleus and not in the anterior part, which coincides with the position of the lens nucleus.

We then extended our computation to realistic exposure scenarios based on the weather data of the five cities. The results suggest that the time course of temperature is variable in different cities, even when the daily peak temperature is identical. Different ambient metrics, yearly and monthly averaged temperature, WBGT, UV radiation, and lens temperature showed modest correlations with NUC prevalence ($0.3 < R^2 < 0.6$, except for UV and lens temperature), along with a $p$-value $> 0.05$. The difference in prevalence in subjects in their 50s and 60s indicated the importance of accounting for the cumulative effect.

We then introduced a new metric, the "cumulative thermal dose in the lens," which is defined as the increase in lens temperature above a certain temperature integrated over a specific period.

This metric showed a higher correlation with the prevalence difference of the lens nucleus cataract from 50s to 60s, as compared to the daily peak lens temperature and UVA radiation ($R^2$=0.920 and $R^2$=0.85, with and without solar radiation, respectively).

In this study, the threshold was set at 37.0 °C. We did not observe a significant difference in correlation when varying the threshold from 36.8 to 37.4 °C. For example, the coefficient of determination for the case with solar radiation changed from 0.96 to 0.72 ($p<0.05$). Our assumption in computation did not consider human activity carefully and thus, fine tuning of the threshold would be arbitrary. It is necessary to set an appropriate threshold from a biological viewpoint (e.g., an in vitro study) in future studies. This metric is considered as a potential exposome in cataractogenesis.

The cumulative thermal dose without solar radiation is relatively low to correlate with the increase in prevalence. However, the lens temperature was higher than 37.0 °C for cases with solar radiation. A similar cumulative thermal dose was also observed in other temperate cities. This indicates that simultaneous exposure to heat and solar radiation (e.g., due to occupational exposure) is needed to cause heat stress in a temperate zone. One weakness of the cumulative thermal dose in the lens as a metric for exposure is that if it is applied to the age range of 0 to 50 years, the estimated prevalence of NUCs may not coincide with the observed value, potentially because of the age dependence of the eye physiology.

Our results provide no evidence to refute the findings that UVA causes nuclear cataract, which was demonstrated in an animal model, wherein guinea pigs were exposed to chronic, low-level UVA light for 4–5 months developed NUCs [68]. It was speculated that most UVB is absorbed within the anterior lens surface, whereas most UVA is absorbed by the lens and reaches the nucleus. Several mechanisms are involved in NUC development, including posttranslational modifications to proteins such as truncation, isomerization, deamidation, methylation, phosphorylation, acetylation, carbamylation, glycation, UV filter modification, and oxidation [69]. Among these, oxidation plays a key role in NUC formation [70]. UVA light mediates the oxidation of ascorbic acid due to the activation of UVA sensitizers in aged human lens water-insoluble proteins. Oxidation products are incorporated into lens proteins by the Maillard reaction, and ascorbate glycation occurs in the human lens nucleus [71]. Glutathione (GSH) protects against covalent binding of UV filters such as kynurenine to crystalline, which would otherwise cause fluorescence in the human lens [72]. However, GSH in the lens nucleus decreases rapidly after the age of 40, leading to age-related susceptibility to sunlight-induced lens damage.

In addition, the cumulative thermal dose causes several age-related changes in the crystalline lens. Exposure of the lens to heat risks and loss of alpha crystallin with conversion into insoluble protein is associated with a progressive increase in lens stiffness and NUC incidence and development [69]. Short-term exposure of lenses to 50 °C induced loss of crystallin, accompanied by an increase in insoluble protein and lens stiffness [73]. Conti et al. reported the use of transgenic techniques to extend the lifespan of mice by decreasing their core body temperature [74]. Thermal interventions avoiding high temperatures may not only reduce the risk of age-related lens changes but also heat-induced denaturation of structural proteins in the whole body.

The temperature increase in the internal tissues (including the lens) is used to set the restriction in guidelines at frequencies below 300 GHz [75], and the corresponding power and energy density are used for practical assessment. Only the incident power density and energy (external quantities) are defined at wavelengths shorter than 1 mm (above 300 GHz). In an electromagnetic field < 300 GHz, no cumulative effect is considered as the exposure level is small. In general, the rise in lens temperature was less than 0.1 °C in the general public. In addition, the safety guidelines for optical radiation and lens damage are considered for intense and brief exposures [76]. Considering the cumulative thermal dose in the lens, this may not violate current international safety guidelines [75].

The limitations and uncertainties of this study are summarized as follows:

i) Validation of lens temperature (see Appendix B) with a computational approach cannot be conducted completely as noninvasive measurement of the temperature inside the eye is not feasible. Our computational method was validated for a rabbit eye with local anesthesia [77, 78]. In addition, the dominant factor was shown to be the blood flow in the choroid and iris in previous studies. However, only a limited number of measurements have been reported to date for accurately evaluating blood flow. Considering the blood flow in the central nervous system, the blood flow in the cortical part of the brain for the elderly has been reported to be half of that in younger adults [79]. However, the main physical factor affecting the lens temperature was found to be via the core temperature rather than the local temperature rise ambient temperature. Thus, this effect would not be significant, though this is one of the uncertainty factors (see Appendix). Some modeling studies have discussed the fluid dynamics of the anterior chamber in the eye (e.g., [80-82]), which was not discussed here. Most considered the heat load directly on the eye rather than the heat transfer mechanism evaluated in this study. Instead, solar radiation with a penetration depth of a few millimeters may result in a complex temperature distribution; power absorption outside the eyeball should not be neglected [41]. In addition, the time scale of temperature is rather long, and other uncertainties would be significant. In addition, the lens temperature considered here is the mid-part of the lens rather than the cortical part, resulting in marginal influence.

ii) The threshold of lens temperature for potentially causing NUC formation was set at 37.0 °C, and sensitivity analysis was conducted. This threshold is based on our computational results, though this is comparable to the $CEM_{43}$ for long-term exposure assumption. Further research on biological aspects is needed to set an appropriate reference value.

iii) The thermoneutral condition was assumed to be 30 °C for all test subjects. However, several uncertainties exist in setting thermoneutral conditions, such as body composition, clothing, energy expenditure, age, and gender. There are only a few studies on the degree of long-term heat acclimatization. Thus, modeling the elderly in the tropics was simply based on the combination of people in the tropical zone and that of elderly subjects in the temperate zone. To the best of our knowledge, no studies have provided data on this aspect. Modeling solar radiation during daily activities is difficult because it depends on individual activity levels. This suggests the necessity of statistical analysis for data acquired through interviews with subjects.

iv) Exercise/activity or increased metabolic rate were not considered in the analysis, and a resting condition was assumed. If we considered modest activity (METs = 4, corresponding to walking for exercise, 3.7 W/kg), a core temperature increase of 0.2–0.3 °C was expected. Additional computation combined with epidemiological studies may be needed in future research.

v) The validation for the model for the elderly in tropical climates proposed here is difficult due to a lack of available data. Our model only extends the parameters related to the decline of sweating due to aging, which is reported in previous studies [83, 84]. The tuning of the parameters is an uncertainty factor.

vi) The application of epidemiological data reported in our previous study as compared with the computational data is another uncertainty factor. The average age of the patients in their 50s was 57.6 years, which is 5.2 years higher than that in Sanya. However, NUC prevalence in Monzen was lower than that in other cities. Thus, no compensation and/or extrapolation in the comparison was applied in this study. In addition, no epidemiological data are available except for five cities. Additional large-scale epidemiological studies are needed to acquire the prevalence and questionnaires on activity are required to estimate exposure to solar radiation.

## 6. Conclusion

With a large-scale integrated computational method of bioheat modeling and thermoregulation, we evaluated the daily lens temperature change in test subjects with different thermoregulatory responses considering climate and age for the first time. We then discussed its correlation with the prevalence of NUCs. Our computational estimation of the cumulative thermal dose in the lens was found to be highly correlated with NUC prevalence than with the remaining parameters (daily peak lens temperature and UV radiation). This is because of the difference in core temperature, which is attributed to the thermoregulatory response. This result indicated that one dominant physical agent, UV exposure, would be different for NUCs compared to cortical cataracts. To enhance applicability to exposure assessment in practical cases, the cumulative WBGT, which corresponds to the cumulative thermal dose in the lens, is proposed to assess NUC prevalence. Computation considering more detailed epidemiological data (e.g., human behavior) is needed to quantify this gap, in addition to the necessity of further epidemiological studies. Our findings contribute to a better understanding of the parameters that are useful for predicting NUC formation worldwide in the future, particularly considering the impacts of climate change expected within the next few decades.

*Appendix A. Parameters used computational code*

The thermal parameters used in the present study were the same as those in [42]. These parameters were listed in Table A. The values of $h_{x,\,m}$ in each body part for the temperate and tropical subjects are same in [23] listed in Table A2.

Table A1. Thermal parameters and mass density of tissues. $K$ is the heat conductivity, $C$ is the specific heat, $B$ is the blood perfusion rate, $\rho$ is the mass density, and $M$ is the basal metabolic rate.

| Tissue type | $K$ [W/m·°C] | $C$ [J/kg·°C] | $B$ [W/m³·°C] | $\rho$ [kg/m³] | $M$ [W/m³] |
|---|---|---|---|---|---|
| Air | 0.03 | 1000 | 0 | 1.20 | 0 |
| Skin | 0.42 | 3600 | 3680 | 1125 | 1620 |
| Muscle | 0.50 | 3800 | 2700 | 1047 | 480 |
| Fat | 0.25 | 3000 | 1700 | 916 | 300 |
| Bone cortical | 0.37 | 3100 | 3400 | 1990 | 610 |
| Bone cancellous | 0.41 | 3200 | 3300 | 1920 | 590 |
| Cartilage | 0.47 | 3600 | 9000 | 1097 | 1600 |
| Nerve | 0.46 | 3400 | 40000 | 1038 | 7100 |
| Bone marrow | 0.22 | 3000 | 32000 | 1040 | 5700 |
| Gray matter | 0.57 | 3800 | 40000 | 1038 | 7100 |
| White matter | 0.50 | 3500 | 40000 | 1038 | 7100 |
| Cerebellum | 0.57 | 3800 | 40000 | 1038 | 7100 |
| CSF | 0.62 | 4000 | 0 | 1007 | 0 |
| Vitreous humor | 0.58 | 4000 | 0 | 1009 | 0 |
| Cornea | 0.52 | 3600 | 0 | 1076 | 0 |
| Lens | 0.40 | 3000 | 0 | 1053 | 0 |
| Retina | 0.58 | 4000 | 0 | 1026 | 0 |
| Sclera | 0.58 | 3800 | 75000 | 1026 | 22000 |
| Heart | 0.54 | 3900 | 54000 | 1030 | 9600 |
| Liver | 0.51 | 3700 | 68000 | 1030 | 12000 |
| Lung(inner) | 0.14 | 3800 | 9500 | 260 | 1700 |
| Lung(outer) | 0.14 | 3800 | 9500 | 1050 | 1700 |
| Kidneys | 0.54 | 4000 | 270000 | 1050 | 48000 |
| Small intestine | 0.57 | 4000 | 71000 | 1043 | 13000 |
| Large intestine | 0.56 | 3700 | 53000 | 1043 | 9500 |
| Gall bladder | 0.47 | 3900 | 9000 | 1030 | 1600 |
| Spleen | 0.54 | 3900 | 82000 | 1054 | 15000 |
| Stomach | 0.53 | 4000 | 29000 | 1050 | 5200 |
| Pancreas | 0.54 | 4000 | 41000 | 1045 | 7300 |
| Blood | 0.56 | 3900 | 0 | 1058 | 0 |
| Blood vessel | 0.56 | 3900 | 9100 | 1040 | 1620 |
| Body fluid | 0.56 | 3900 | 0 | 1010 | 0 |
| Bile | 0.55 | 4100 | 0 | 1010 | 0 |
| Lymph | 0.56 | 3900 | 0 | 1040 | 0 |
| Glands | 0.53 | 3500 | 360000 | 1050 | 64000 |
| Bladder | 0.43 | 3200 | 9000 | 1030 | 160 |
| Testicles | 0.56 | 3900 | 360000 | 1044 | 64000 |
| Tooth | 0.37 | 3100 | 3400 | 2160 | 590 |
| Ligaments | 0.50 | 3600 | 9000 | 1220 | 1600 |
| Mucous membrane | 0.50 | 3600 | 9000 | 1040 | 1600 |
| Nails | 0.50 | 3600 | 0 | 1030 | 0 |
| Dura | 0.50 | 3600 | 9100 | 1125 | 0 |
| Tongue | 0.54 | 3800 | 2700 | 1047 | 480 |
| Skull | 0.39 | 3100 | 3300 | 1850 | 610 |
| Trachea | 0.47 | 3650 | 3700 | 1100 | 1600 |
| Testis prostate | 0.53 | 3800 | 65000 | 1050 | 64000 |
| Tendon | 0.50 | 3500 | 6300 | 1100 | 1600 |
| Hypothalamus | 0.57 | 3800 | 40000 | 1038 | 7100 |
| Duodenum | 0.53 | 4000 | 29000 | 1050 | 5200 |
| Esophagus | 0.53 | 4000 | 29000 | 1050 | 5200 |
| Anterior chamber | 0.58 | 4000 | 0 | 1009 | 0 |
| Iris | 0.52 | 3600 | 35000 | 1026 | 10000 |
| Sclera | 0.58 | 3800 | 150000 | 1026 | 40000 |

Table A2. Blood volume $V$ and counter-current coefficients of temperate and tropical subjects in each body part.

|  | Volume ($V$) [m³] | Counter-current coefficients $h_x$ | |
| --- | --- | --- | --- |
|  |  | Temperate subjects | Tropical subjects |
| Head and body | 0.0374 | 0.0 | 0.0 |
| Upper arms | 0.0019 | 0.7 | 3.1 |
| Lower arms | 0.0010 | 1.9 | 6.2 |
| Hands | 0.0005 | 0.6 | 2.6 |
| Upper legs | 0.0066 | 25.0 | 55.0 |
| Lower legs | 0.0024 | 33.0 | 70.7 |
| Feet | 0.0011 | 23.0 | 110.0 |

*Appendix B. Effect of Parameters for Temperature in Human Eye*

The first exposure scenario was the validation of the corneal temperature at different ambient temperatures. In this scenario, clothing was adjusted to prevent the core temperature from changing due to ambient conditions. A comparison was made for ambient temperatures ranging from 18 and 27 °C, which are based on the data in [85, 86]. Note that noninvasive measurement of the lens temperature is not feasible.

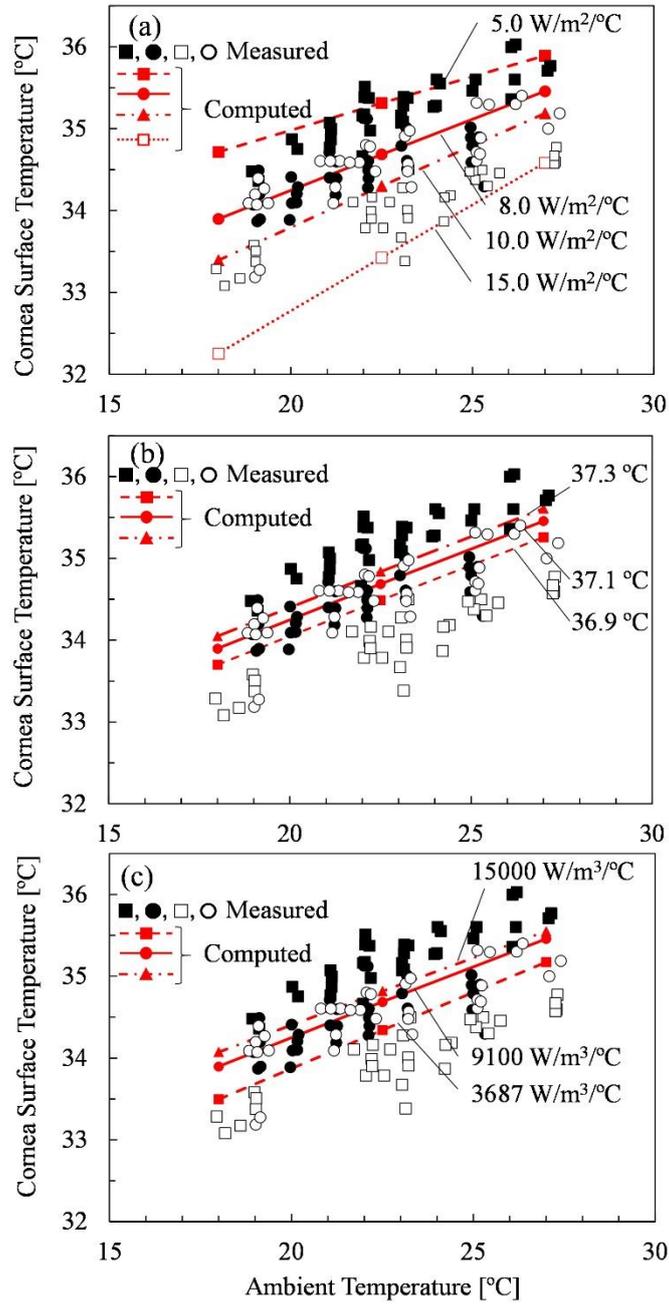

Figure B1. Corneal temperature at different ambient temperatures. Measured data was reported in [55]. Computed data was changed the parameter of (a) heat transfer rate between cornea and air, (b) core temperature, and (c) skin blood flow.

According to previous studies, there are three dominant factors affecting local temperature and temperature due to heat load: heat transfer rate between air and cornea, skin blood flow, and core temperature. Note that blood flow in the skin and iris may be dominant for the corneal temperature as the heat diffusion length is of the order of 10 mm.

First, the heat transfer rate on the corneal temperature was computed for different ambient temperatures and then compared with the measured value. As shown in Fig. A1 (a), the heat transfer rate may influence the slope of the regression line characterizing the ambient and corneal temperatures. As the slope for subjects in the previous report varies in the range of 0.13–0.19, the heat transfer coefficient should be in the range of 5–7 W/m$^2$/°C. Note that considering the surface area difference in the voxel and smoothed models, the actual value was approximately 7–10 W/m$^2$/°C.

Then, we derived the thermal parameters in whole-body models by comparing the corneal temperatures reported in previous studies. The heat transfer coefficient between the cornea and air was 7–10 °C, which is 2–3 times lower than those reported previously for rabbits [78]. One of the possible reasons for this could be that blinking of the eye is much slower in rabbits than in humans, and thus the derived coefficient may include the effect. When the eye is closed, the cornea is warmed by the skin blood, which has a higher temperature than that of the air. Thus, the above-mentioned value of the heat transfer coefficient can be considered as a combination of blinking and evaporative heat loss from the eye surface, in addition to radiation, similar to the skin. This suggested that the core temperature serves as an intercept of the regression line characterizing the corneal temperature.

The blood flow in the skin may influence the basal temperature and approximately coincides with the measured value for blood flow of 9,100–4000 W/m$^3$/°C. A similar effect was observed for the core temperature. For simplicity, we set this value at 37.0 °C. Although not shown here, the impact of iris blood flow was marginal at a relatively longer distance. Additionally, the iris blood flow was sufficiently high.